\documentclass[aps,prd,twocolumn,nofootinbib,longbibliography,10pt]{revtex4-2}
\usepackage{etoolbox}
\usepackage{dcolumn,tensor,nicefrac}
\usepackage{amsmath,amssymb,amsfonts,mathtools}
\usepackage{mathrsfs,bbold}
\usepackage{graphicx}
\usepackage[colorlinks=true,urlcolor=blue,citecolor=red,linkcolor=blue]{hyperref}
\usepackage{accents}
\newlength{\dhatheight}
\usepackage{natbib}
\usepackage{wrapfig}
\usepackage{flushend,BOONDOX-cal,BOONDOX-frak}

\makeatletter
\newsavebox{\@brx}
\newcommand{\llangle}[1][]{\savebox{\@brx}{\(\m@th{#1\langle}\)}%
  \mathopen{\copy\@brx\kern-0.5\wd\@brx\usebox{\@brx}}}
\newcommand{\rrangle}[1][]{\savebox{\@brx}{\(\m@th{#1\rangle}\)}%
  \mathclose{\copy\@brx\kern-0.5\wd\@brx\usebox{\@brx}}}
\makeatother
\begin{document}
\title{\textbf{Area spectrum and black hole thermodynamics}}
\author{Arpita Jana}
\email{janaarpita2001@gmail.com}
\affiliation{Department of Astrophysics and High Energy Physics, S. N. Bose National Centre for Basic Sciences, JD Block, Sector-III, Salt Lake City, Kolkata-700 106, India}
\author{Manjari Dutta}
\email{chandromouli15@gmail.com}
\affiliation{Department of Astrophysics and High Energy Physics, S. N. Bose National Centre for Basic Sciences, JD Block, Sector-III, Salt Lake City, Kolkata-700 106, India}
\author{Sunandan Gangopadhyay}
\email{sunandan.gangopadhyay@gmail.com}
\affiliation{Department of Astrophysics and High Energy Physics, S. N. Bose National Centre for Basic Sciences, JD Block, Sector-III, Salt Lake City, Kolkata-700 106, India}
\begin{abstract}
\noindent The role of horizon area quantization on black hole thermodynamics is investigated in this article. The coefficient appearing in the quantization of area is fixed by an appeal to the saturated form of the Landauer's principle. Then by considering transition between discrete states of the event horizon area which in turn is equivalent to transitions between discrete mass states of the black hole, the change in the mass can be obtained. The change in mass is then equated to the product of the Hawking temperature and change in entropy of the black hole between two consecutive discrete states applying the first law of black hole thermodynamics. This gives the corrected Hawking temperature. In particular, we apply this technique to the Schwarzschild black hole, the quantum corrected Schwarzschild black hole, the Reissner-Nordstr\"{o}m black hole which is a charged black hole, and the rotating Kerr black hole geometry, and obtain the corrected Hawking temperature in each of these cases. We then take a step forward by inserting this corrected Hawking temperature in the first law of black hole thermodynamics once again to calculate the entropy of the black hole in terms of the horizon area of the black hole. This leads to logarithmic and inverse corrections to the entropy of the black hole. \end{abstract}
\maketitle
\section{Introduction}
\noindent After the classic development of Einstein's theory of general relativity \cite{Einstein1,Einstein2}, Bekenstein and Hawking made remarkable developments \cite{Hawking1,Hawking2,Bekenstein1,Bekenstein2} to make connections between general relativity and thermodynamics. Hawking's pioneering research was based on the radiation emitted from black holes, which can be characterized by a temperature known as the Hawking temperature. This in turn also leads to the famous Hawking area law formula for the entropy of the black hole. For a thorough discussion, the reader is referred to \cite{Wald}. However, the derivation of this effect considers a quantum field in a classical black hole background and thus the phenomenon of Hawking radiation arises, which is purely a quantum mechanical effect. The analysis is semi-classical in nature for obvious reasons mentioned above. The result is not only remarkable but it also opens up very important question, namely, the effects of quantization of gravity \cite{Kiefer, Rovelli, Carlip, AmatiCiafaloni, KonishiPaffuti}. Quantizing gravity is itself a very difficult problem and has been a holy grail of theoretical physics for the last four decades. In the context of black holes, the effects of quantizing gravity would mean quantum fluctuations of the event horizon of the black hole. This would in turn mean a modification of the nature of Hawking radiation and the entropy of the black hole as has been rightly pointed out in \cite{BekensteinMukhanov}. The effects would obviously be prominent in case of black holes whose sizes are comparable to the Planck length, but they would in principle still be there for those black holes which are quite large in size compared to the Planck scale.
\par \noindent Long ago, Bekenstein in his Ph.D thesis (1972) \cite{Bekenstein1}, first introduced the idea that the surface area of the black hole can be quantized in units of Planck length squared. Since the area and the mass of the black hole are related with each other, as $A=16\pi \left(\frac{G_{0}^2M^2}{c^{4}}\right)$\footnote{$G_{0}$ is the Newton's gravitational constant, $M$ is the mass of the black hole and $c$ is the speed of light in vacuum.} for the Schwarzschild black hole, mass of the black hole should also be quantized \cite{BekensteinMukhanov}. This is referred to as the mass spectrum of the black hole in \cite{BekensteinMukhanov}. That particularly means that in Planckian length scale, there are discrete energy states of a black hole. Corresponding to these discrete energy levels, there is entropy and hence for the transition between two consecutive levels, there is a change in entropy of the black hole. This would also mean that there can be information loss.  
This is based on the concept of Hawking's area theorem which states that the area of a black hole cannot decrease. It should be noted however that if higher order curvature corrections in general theory of relativity are considered, then the entropy is no longer proportional to the area of the black hole, rather there will be some corrections. These include mostly power-law corrections in entropy, but there are some quantum gravitational methods where the black hole entropy contains subleading logarithmic corrections. This corrected relation between the area and the entropy gives rise to the non-uniform quantization of black hole area. Some of the important impacts of this non-uniform area spectrum on gravitational waves and black hole echos are discussed in \cite{KRS1,KRS2}.
\noindent In this paper, we start by revisiting the connection between black hole thermodynamics and the Landauer's principle in information theory \cite{Landauer}. The Landauer's principle was first introduced in 1961 and relates the removal of one bit of information to the amount of energy loss of the operating system governed by its temperature. In the context of black hole thermodynamics, we propose that the change in energy of the black hole governed by the first law of black hole thermodynamics is equal to the Landauer loss in energy related to the erasure of one bit of information. This gives a bound on a parameter related to the change in area of the black hole as we shall find in the subsequent discussion.\\
\noindent We then proceed to consider the different black holes one by one and start from their area quantization. Since the area of the black hole is related to its mass, we get the mass spectrum of the black hole which is also found to be quantized.\\
\noindent Considering transitions between consecutive mass states of the black hole, leads to the corrected Hawking temperature of the black hole. This is obtained from the first law of black hole thermodynamics once again.\\
\noindent We then take the approach in \cite{BekensteinMukhanov} and also follow the analysis taken recently in \cite{Bagchi}. Considering the fourth order of inverse mass of the Schwarzschild black hole, we obtain the mass spectrum which leads to the corrected Hawking temperature for the Schwarzschild black hole. Using the corrected Hawking temperature, we apply the first law of black hole thermodynamics once again to calculate the corrected entropy of the Schwarzschild black hole. In section \ref{III}, we introduce the mass of the quantum corrected Schwarzschild black hole and calculate the correction of Hawking temperature and entropy for the black hole. In section \ref{IV} and section \ref{V}, we consider the Reissner-Nordstr{\"o}m and Kerr black hole geometries respectively and apply the same analysis. In the final section, we discuss the results of our analysis.

\section{Area quantization of Schwarzschild black hole and Landauer's principle}\label{II}
\noindent Landauer, in 1961, in one of his revolutionary works \cite{Landauer}, first attempted to unify thermodynamics and digital computing. Typically, the Landauer's principle states \cite{Bennett} that all data processing operations require the dissipation of at least $k_{B}T\ln{2}$ amount of energy to erase one bit of information. Mathematically this implies
\begin{equation}
    \Delta E_{L}\ge k_{B}T\ln{2}~.
\end{equation}
where $k_{B}$ is the Boltzmann's constant and L in the subscript corresponds to Landauer's energy. In a very recent work \cite{CortesLiddle}, the Landauer's principle was applied in the background of Schwarzschild geometry and it was concluded that the energy dissipated to the surroundings is exactly the same to that of the Landauer's energy corresponding to Hawking temperature $T_{H}$. Following this remarkable observation, we shall apply this in the context of area quantization of black holes.
\noindent  In the second half of twentieth century, heuristic theory about the quantization of black hole area was developed, and the area spectrum\footnote{The spectrum of the area of the black hole resembles the energy spectrum of quantum harmonic oscillator; hence, in order to introduce the zero-point effect, we have taken $(n+\frac{1}{2})$ instead of n.} in semi-classical approximation was given in natural units  ($\hbar = c=k_{B}=1$) as \cite{BekensteinMukhanov,Bagchi}
\begin{equation}\label{areaspectrum}
    A(n)=\alpha \left(n+\frac{1}{2}\right)l_{P}^{2}~
\end{equation}
where $\alpha$ is a constant, $n$ is positive integer\footnote{In the semi-classical limit, $n\gg 1$.} and $l_{P}$ is Planck's length\footnote{Planck's length $l_{P}$ is given as $\sqrt{\frac{G\hbar}{c^{3}}}$ with the constants carrying their usual meaning. In natural units, $G \sim l_{P}^{2} \sim \frac{1}{m_{P}^{2}}$ ; where $m_{P}$ is the Planck's mass.}. This actually agrees with the concept of area quantization in canonical quantum gravity \cite{Ashtekar}. 
Using the famous area by four law for the entropy of a black hole \cite{Hawking1,Hawking2}, one can obtain the entropy spectrum as
\begin{equation}\label{entropyspectrum}
    S_{\text{BH}}\simeq \frac{A}{4G_{0}}\simeq \frac{\alpha (n+\frac{1}{2})}{4}~.
\end{equation}
 Now the thermal energy dissipated corresponding to this entropy is given from the first law of thermodynamics as
\begin{equation}
    \Delta E_{\text{BH}}=T_{H}\Delta S_{\text{BH}}\simeq T_{H}\frac{\alpha}{4}~.
\end{equation}
Here, $T_{H}$ is the Hawking temperature for Schwarzschild black hole; expressed as 
\begin{equation}\label{Hawkingtemp}
T_{H}=\frac{m_{P}^{2}}{8\pi M}~.
\end{equation}
Taking the clue from the work in \cite{CortesLiddle}, we propose that this energy dissipated due to Hawking radiation is sufficient to remove one bit of information from the semiclassical black holes, that is, we propose that the above change in energy of the black hole is equal to the change in Landauer's energy
\begin{equation}
    \Delta E_{BH}=\Delta E_{L}~.
\end{equation}
This relation fixes the value of the semi-classical parameter $\alpha$ to be 
\begin{equation}
\alpha \ge 4\ln{2}~.
\end{equation}
As discussed in \cite{BekensteinMukhanov}, from information theory point of view, the minimum value of $\alpha=4\ln{2}$ is the one which makes the difference of entropy between two consecutive energy levels exactly one bit. \\
\noindent Now we shall move to find the mass spectrum of the Schwarzschild black hole. The well known area expression for the Schwarschild black hole is $A=4\pi r_{s}^{2}\simeq 16\pi G_{0}^{2}M^{2}$; where $r_{s}\simeq 2G_{0}M$ is the Schwarzschild radius in natural units; related to Newton's gravitational constant $G_{0}$ and mass of the black hole $M$. Due to the relation between the area and the mass of the black hole given in the above relation, there should exist discrete mass states that is one can quantize the mass of the black hole also. So, the mass spectrum for the Schwarzschild black hole can be written as \cite{Bagchi}
\begin{equation}\label{massspectrum1}
    M(n) \simeq \frac{\sqrt{A}m_{P}^{2}}{4\sqrt{\pi}}\simeq \frac{m_{P}}{4}\sqrt{\frac{\alpha(n+\frac{1}{2})}{\pi}}~.
\end{equation}
This denotes the mass of the black hole in the $n^{th}$ energy state. If the black hole transition occurs from $(n+1)^{th}$ state to $n^{th}$ state, the mass difference will take the form
\begin{equation}\label{massdiff1}
    \Delta M\simeq \frac{m_{P}}{8}\sqrt{\frac{\alpha}{\pi n}}\left(1-\frac{1}{2n}+\frac{13}{32n^{2}}\right)~.
\end{equation}
Substituting the value of $n$ from eq.(\ref{massspectrum1})in terms of $M$, one can recast eq.(\ref{massdiff1}) as
\begin{equation}\label{diffmass}
    \Delta M\simeq \frac{\alpha m_{P}^{2}}{32\pi M}\left(1-\frac{\alpha m_{P}^{2}}{64\pi M^{2}}+\frac{\alpha^{2}m_{P}^{4}}{2\times (32)^{2}\pi^{2}M^{4}}\right)~.
\end{equation}
Note that the above result is slightly different from the one presented in \cite{Bagchi}. The reason for this difference is due to the fact that a correction of  
 $\mathcal{O}\left(\frac{1}{M^{2}}\right)$ gets generated from the next higher order term. By taking care of this, the correct coefficient of the $\mathcal{O}\left(\frac{1}{M^{2}}\right)$ term is obtained in eq.(\ref{diffmass}). This is in contrast to the result obtained in \cite{Bagchi}.\\
\noindent This mass difference in eq.(\ref{diffmass}) introduces a correction in the black hole temperature following the first law of black hole thermodynamics
\begin{align}
    \mathcal{T}=&\frac{\Delta M}{\Delta S_{BH}}\nonumber \\ \simeq &\frac{ m_{P}^{2}}{8\pi M}\left(1-\frac{\alpha m_{P}^{2}}{64\pi M^{2}}+\frac{\alpha^{2}m_{P}^{4}}{2\times (32)^{2}\pi^{2}M^{4}}\right)\nonumber \\ 
    =& T_{H}\left(1-\frac{\alpha m_{P}^{2}}{64\pi M^{2}}+\frac{\alpha^{2}m_{P}^{4}}{2\times (32)^{2}\pi^{2}M^{4}}\right)~.
\end{align}
Here, the leading term is the semi-classical Hawking temperature $T_{H}$ mentioned in eq.(\ref{Hawkingtemp}), and it is corrected upto $\mathcal{O}(\alpha^{2})$.
 Now we shall use this corrected Hawking temperature and apply the first law of black hole thermodynamics once again. This leads to (upto $\mathcal{O}(\frac{1}{M^{2}})$)
\begin{align}\label{bhentropy}
    \mathcal{S}_{BH} =&\int \frac{dM}{\mathcal{T}} \nonumber \\ \simeq  &\frac{4\pi}{m_{P}^{2}}\left [ M^{2}+\frac{\alpha m_{P}^{2}}{32\pi}\ln{\left(\frac{M}{m_{P}}\right)} + \frac{\alpha^{2}m_{P}^{4}}{4\times(32)^{2}\pi^{2}M^{2}}\right ]~.
\end{align}
The above equation displays the relation between the mass and the entropy of the black hole, but to compare it with the Hawking entropy, we have to rewrite it in terms of the area of the black hole. Inserting eq.(\ref{massspectrum1}) into eq.(\ref{bhentropy}), one can recast the above equation as
    \begin{align}
         \mathcal{S}_{BH} \simeq \frac{Am_{P}^{2}}{4}+\frac{\alpha}{16}\ln{A}+\frac{\alpha^{2}}{64Am_{P}^{2}} +\frac{\alpha}{8}\ln{\left(\frac{m_{P}}{4\sqrt{\pi}}\right)}~.
    \end{align}
The leading order correction in entropy formula is logarithmic, while the subleading term is proportional to the inverse of the black hole area. We shall now carry out a similar analysis for different kinds of black hole geometries in the following sections.

\section{Mass spectrum, corrected temperature and entropy of quantum improved Schwarzschild black hole}\label{III}
\noindent Quantum theory of gravity is one of the most elegant theories needed to be developed ever to unify quantum mechanics and general relativity. One of the most achievable schemes to do this is the renormalization group \cite{Reuter1,ReuterSaueressig,Percacci} approach where the asymptotic safety scenario has been taken care of. In the beginning of the $21^{st}$ century, Bonanno and Reuter, in their classic work \cite{BonannoReuter}, first applied this scheme to Schwarzschild black hole geometry; where the Newton's gravitational constant was replaced by the renormalization group flow of gravitational constant. These types of quantum mechanically corrected black holes are termed as ``\textit{quantum corrected (improved) black holes}". 
\noindent The general scheme to obtain a quantum improved geometry is explicitly discussed in \cite{BonannoReuter, PawlowskiStock, Platania, ReuterWeyer}. The first step is to consider the renormalization group flow equation and obtain the form of coupling constants in terms of the cut-off momentum scale, while the second step leads to define the momentum cut-off scale in terms of the radial coordinates in spherically symmetric geometries. There are mainly two ways to identify the cut-off scale; one of which is to construct invariant curvature scalars like the Kretschmann scalar and the Ricci scalar and the other method is based on the curves towards the UV fixed limit point. Now, to finally express the quantum improved geometries, there are three ways to follow. Firstly, one can directly impose the flowing coupling constants into the classical solution. The second approach can be to put the same in the equations of motion; e.g. one can put the running gravitational constant and cosmological constant to get the quantum improved Einstein equations
\begin{equation}
    G_{\mu \nu} = 8\pi G(x) T_{\mu \nu}-\Lambda(x)g_{\mu \nu}
\end{equation}
\noindent where the flow of matter coupling was ignored. The third and the final approach can be improving the action; where one can put the position dependent couplings in the Lagrangian of the system. The quantum improved Einstein-Hilbert action has the form of 
\begin{equation}
    \mathcal{S}_{EH}=\frac{1}{16\pi}\int \frac{d^{4}x}{G(x)}\left(R-2\Lambda(x) \right)~.
\end{equation}
In our analysis, everything is executed by setting $\Lambda = 0$. We followed the first approach to obtain the quantum improved geometry by replacing the running couplings in the solution level. The motivation for considering the simplest approach has been explicitly discussed in \cite{Ishibashi,JanaSenGangopadhyay}.
After the quick revisit of quantum improved geometry, we shall now move to our analysis.\\
\noindent For a general spherically symmetric black hole, one can can write the metric in the form of 
\begin{equation}\label{metric}
    ds^{2}=-f(r)dt^{2}+f(r)^{-1}dr^{2}+r^{2}(d\theta^{2}+\sin^{2}{\theta}d\phi^{2})
\end{equation}
where $f(r)$ is the lapse function of the black hole and in the case of the quantum corrected Schwarzschild black hole, the form of the lapse function reads
\begin{equation}
    f(r)\simeq 1-\frac{2G(r)M}{r}
\end{equation}
where $M$ is the mass of the black hole and the running gravitational constant $G(r)$ satisfies the renormalization group flow equation \cite{HarstReuter,EichhornVersteegen}
\begin{equation}
    k\frac{d\tilde{G}(k)}{dk} = 2\tilde{G}(k)\left( 1-\frac{\tilde{G}(k)}{4\pi \tilde{\alpha}}\right)
\end{equation}
where $\tilde{G}(k)$ is the dimensionless Newton's constant, defined as $\tilde{G}(k)=k^{2}G(k)$ and $\tilde{\alpha}$ is a numerical constant corresponding to a fixed point $\tilde{G}_{\star}=4\pi \tilde{\alpha}$
and flows with the momentum cut-off scale as \cite{BonannoReuter}
\begin{equation}
    G(r)= \frac{G_{0}}{1+\frac{\tilde{\omega}G_{0}}{r^{2}}}
\end{equation}
where $\tilde{\omega}$ is a small positive constant related to quantum gravity correction, $G_{0}$ is the usual Newton's gravitational constant and $r$ is the radial distance from the singularity ($r=0$) of the black hole and the relation between the radial distance and the momentum cut-off scale is discussed in \cite{HarstReuter,EichhornVersteegen}.
\noindent The lapse function of this kind of quantum corrected non-rotating neutral black hole is written in natural units as \cite{BonannoReuter}
\begin{equation}
    f(r) = 1-\frac{2G(r)M}{r} = 1-\frac{2G_{0}Mr}{(r^{2}+\tilde{\omega}G_{0})}~.
\end{equation}
\noindent To find the event horizon of the black hole, we shall put $f(r)=0$ and the event horizon can be obtained upto $\mathcal{O}(\tilde{\omega}^{2})$ as \cite{qsch}
\begin{align}
    r_{h}= & G_{0}M+\sqrt{G_{0}^{2}M^{2}-\tilde{\omega}G_{0}}\nonumber \\
     \simeq & \frac{2M}{m_{P}^{2}}\left(1-\frac{\tilde{\omega}m_{P}^{2}}{4M^{2}}-\frac{\tilde{\omega}^{2}m_{P}^{4}}{16M^{4}} \right ) + \mathcal{O}(\tilde{\omega}^{3})~.
\end{align}
Now, one can write the area of the quantum corrected Schwarzschild black hole as
\begin{align}
    A= 4\pi r_{h}^{2} \simeq 16\pi M^{2}l_{P}^{4}\left(1-\frac{\tilde{\omega}}{2M^{2}l_{P}^{2}}-\frac{\tilde{\omega}^{2}}{16M^{4}l_{P}^{4}}\right) + \mathcal{O}(\tilde{\omega}^{3})
\end{align}
where $l_{P}$ is the Planck's length.
Now, equating with the area spectrum given in eq.(\ref{areaspectrum}), one can obtain the mass spectrum of the black hole to be
\begin{align}\label{massspectrum}
    M (n) \simeq &\frac{m_{P}^{2}}{4}\sqrt{\frac{A}{\pi}}\left ( 1+ \frac{4\pi \tilde{\omega}m_{P}^{2}}{A}\right ) \nonumber\\
    = & \frac{m_{P}}{4}\sqrt{\frac{\alpha (n+\frac{1}{2})}{\pi}} \left (1+\frac{4\pi \tilde{\omega}}{\alpha (n+\frac{1}{2})}\right ).
\end{align}
Now in the large $n$ limit, that is, $n\gg 1$, one can find the difference between mass for the transition between two discrete states $|n+1\rangle$ and $|n\rangle$ as\footnote{We have restricted our analysis upto $\mathcal{O}(\frac{1}{M^{4}})$; and hence, to calculate the mass difference, in the first term, we have considered upto $\mathcal{O}(\frac{1}{n^{2}})$ term and in the second term, we took upto $\mathcal{O}(\frac{1}{n})$.}
\begin{equation}
\begin{split}\label{massdiff}
    \Delta M &\simeq \frac{m_{P}}{8}\sqrt{\frac{\alpha}{\pi n}}\left (1-\frac{1}{2n}+\frac{13}{32n^{2}}\right ) \\ & ~ ~~~~~~ +\frac{m_{P}\tilde{\omega}}{2n}\sqrt{\frac{\pi}{\alpha n}}\left(-1+\frac{3}{2n} \right ).
\end{split}
\end{equation}
From eq.(\ref{massspectrum}), the expression of $n$ is found to be
\begin{equation}\label{nexpress}
    n\simeq \frac{16\pi M^{2}}{\alpha m_{P}^{2}}\left ( 1 - \frac{\alpha m_{P}^{2}}{32\pi M^{2}}-\frac{\tilde{\omega}m_{P}^{2}}{2M^{2}}-\frac{\tilde{\omega}^{2}m_{P}^{4}}{16M^{4}}\right ).
\end{equation}
Now substituting eq.(\ref{nexpress}) into eq.(\ref{massdiff}), we obtain
\begin{equation}
    \Delta M \simeq \frac{\alpha m_{P}^{2}}{32\pi M}\left(1-\frac{\alpha m_{P}^{2}}{64\pi M^{2}}+\frac{\alpha^{2}m_{P}^{4}}{2\times (32)^{2}\pi^{2}M^{4}}-\frac{\tilde{\omega}^{2}m_{P}^{4}}{16M^{4}}\right).
\end{equation}
It is interesting to observe that $\Delta M$ is independent of $\tilde{\omega}$ but it contains the $\mathcal{O}\left(\tilde{\omega}^{2}\right)$ corrections. This is an important observation of our analysis. As discussed in \cite{ Chamseddine}, for noncommutative spacetime geometry, there does not exist any linear order correction of noncommutative parameter $\theta$ in Einstein-Hilbert action, the higher order corrections start from $\mathcal{O}(\theta^{2})$. This is a surprising similarity between the two theories. \\
 \noindent Now we will see how the temperature of the quantum-corrected Schwarzschild black hole is being corrected due to this mass quantization. From the first law of black hole thermodynamics, the corrected temperature reads
\begin{align}\label{temperature}
    \mathcal{T} = &\frac{\Delta M}{\Delta S} \nonumber \\
    \simeq &\frac{m_{P}^{2}}{8\pi M}\left(1-\frac{\alpha m_{P}^{2}}{64\pi M^{2}}+\frac{\alpha^{2}m_{P}^{4}}{2\times (32)^{2}\pi^{2}M^{4}}-\frac{\tilde{\omega}^{2}m_{P}^{4}}{16M^{4}}\right ).
\end{align}
This is the corrected temperature corresponding to the change in mass and entropy between two successive energy levels. Now, the Hawking temperature of the quantum corrected non-rotating neutral black hole is given by 
\begin{align}\label{Hawkingtemp2}
    T^{QC}_{H}=&\frac{1}{4\pi}f^{\prime}(r_{h})\nonumber \\
    \simeq & \frac{m_{P}^{2}}{8\pi M}\left(1-\frac{\tilde{\omega}m_{P}^{2}}{4M^{2}}-\frac{\tilde{\omega}^{2}m_{P}^{4}}{8M^{4}} \right)
\end{align}
where $f^{\prime}(r_{h})$ denotes the first derivative of the lapse function with respect to $r$; evaluated at the event horizon. If one compares eq.(\ref{temperature}) and eq.(\ref{Hawkingtemp2}), the relation between the renormalization group correction parameter $\tilde{\omega}$ and the quantization parameter $\alpha$ comes out as 
\begin{equation}\label{omegaalpharelation}
    \tilde{\omega}=\frac{\alpha}{16\pi}\ge \frac{\ln{2}}{4\pi}.
\end{equation}
Hence we can write down the eq.(\ref{temperature}) in terms of the Hawking temperature of the renormalization group corrected Schwarzschild black hole mentioned in eq.(\ref{Hawkingtemp2}). The eq.(\ref{temperature}) therefore takes the form
\begin{align}
    \mathcal{T}\simeq T^{QC}_{H}\biggl[ 1 & - \frac{\alpha m_{P}^{2}}{64\pi M^{2}}+\frac{\alpha^{2}m_{P}^{4}}{2\times (32)^{2}\pi^{2}M^{4}}+\frac{\tilde{\omega}m_{P}^{2}}{4M^{2}} \nonumber \\ & -\frac{\alpha \tilde{\omega}m_{P}^{4}}{(8\times 32)\pi M^{4}}+\frac{\tilde{\omega}^{2}m_{P}^{4}}{8M^{4}}\biggr]~.
\end{align}
 Now we shall compute the entropy of the black hole due to the semi-classically corrected temperature eq.(\ref{temperature}). So, again from first law of thermodynamics, one can write 
\begin{equation}
    \mathcal{S} = \int \frac{dM}{\mathcal{T}}.
\end{equation}
Substituting eq.(\ref{temperature}), one can obtain the entropy in terms of the mass of the black hole upto $\mathcal{O}(\tilde{\omega}^{2})$ as\footnote{To make the logarithmic term dimensionless, we have computed the integral with respect to $\frac{M}{m_{P}}$.}
\begin{align}
     \mathcal{S} \simeq \frac{4\pi}{m_{P}^{2}}&\left [ M^{2}+\frac{\alpha m_{P}^{2}}{32\pi}\ln{\left(\frac{M}{m_{P}}\right)}
     \right. \nonumber \\ & \left.+ \frac{\alpha^{2}m_{P}^{4}}{4\times(32)^{2}\pi^{2}M^{2}}-\frac{\tilde{\omega}^{2}m_{P}^{4}}{16M^{2}}\right ].
\end{align}
Now we shall write this entropy in terms of the area of the black hole. So, from eq.(\ref{massspectrum}), we shall now substitute mass in terms of area and by doing so, we obtain the following result for the entropy 
\begin{align}
    \mathcal{S} & \simeq \frac{Am_{P}^{2}}{4}+\frac{\alpha}{16}\ln{A}+\frac{\pi \alpha}{2Am_{P}^{2}}\left(\frac{\alpha}{32\pi}+\tilde{\omega}\right)\nonumber \\ & -\frac{\pi \alpha \tilde{\omega}}{A^{2}m_{P}^{4}}\left( \frac{\alpha}{8}+\pi\tilde{\omega}\right)+\frac{3\alpha^{2}\pi^{2}\tilde{\omega}^{2}}{4A^{3}m_{P}^{6}}+\frac{\alpha}{8}\ln{\left(\frac{m_{P}}{4\sqrt{\pi}}\right)}+2\pi \tilde{\omega}~.
\end{align}
From this expression, we can see that the leading term in the entropy expression follows the usual \textit{``area divided by four"} law and the second leading order term is logarithmic, followed by the other corrections of $\mathcal{O}(\frac{1}{A})$, $\mathcal{O}(\frac{1}{A^{2}})$ and $\mathcal{O}(\frac{1}{A^{3}})$.\\
\begin{figure}[hb!]
\begin{center}
\includegraphics[scale=0.6]{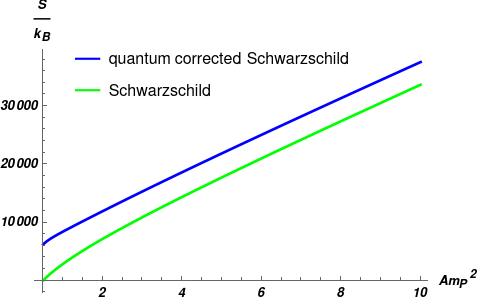}
\caption{Entropy versus area plot for the Schwarzschild and the quantum corrected Schwarzschild black holes\label{entropy}.}
\end{center}
\end{figure}
\noindent In Fig.(\ref{entropy}), we have plotted dimensionless quantities $\frac{S}{k_{B}}$ and $Am_{P}^{2}$ for the Schwarschild and the quantum corrected Schwarzschild black holes. We have put $\tilde{\omega}=0.05$ which is consistent with the eq.(\ref{omegaalpharelation}). We can see from the plot that the corrected entropy for the quantum corrected case is higher than the Schwarzschild black hole case.

\section{Mass spectrum, corrected temperature and entropy of Reissner-Nordstr{\"o}m black hole}\label{IV}
\noindent Reissner-Nordstr$\ddot{o}$m black holes are charged spherically symmetric black holes. These are not relevant to the astrophysical black holes as highly-charged black holes are quickly neutralized by its interactions with surrounding matters \cite{Carroll}. But they have important impacts on the general theoretical analyses. Since, the spherical symmetry is present in this case, the general metric of non-rotating charged black hole looks like eq.(\ref{metric}) with the lapse function given by
\begin{equation}
    f(r) = 1-\frac{2G_{0}M}{r}+\frac{e^{2}G_{0}}{r^{2}}
\end{equation}
where $e$ denotes the charge and $M$ is the mass of the black hole. 
To find the area of the event horizon, we need to find the horizon radius. For charged black hole, radius of the event horizon has the form
\begin{equation}
    r_{h}=G_{0}M+G_{0}M\sqrt{1-\frac{e^{2}}{G_{0}M^{2}}}~.
\end{equation}
Now equating the area of this black hole and semi-classical quantization form of area, one can find the mass spectrum for this black hole to be
\begin{align}\label{massspectrum2}
    M(n)\simeq &\frac{m_{P}^{2}}{4}\sqrt{\frac{A}{\pi}}\left(1+\frac{4\pi e^{2}}{Am_{P}^{2}}\right)\nonumber \\
    = &\frac{m_{P}}{4}\sqrt{\frac{\alpha (n+\frac{1}{2})}{\pi}}+e^{2}m_{P}\sqrt{\frac{\pi}{\alpha(n+\frac{1}{2})}}
\end{align}
where $A$ is the area of the charged black hole. The dependence of $M$ on $n$ is similar to the quantum corrected Schwarzschild case. Now if transition from the state $|n+1\rangle$ to the state $|n\rangle$ occurs, then the mass difference between two discrete energy levels is 
\begin{equation}\label{massdiff2}
\begin{split}
    \Delta M &\simeq \frac{m_{P}}{8}\sqrt{\frac{\alpha}{\pi n}}\left(1-\frac{1}{2n} + \frac{13}{32n^{2}}\right)\nonumber \\ & +\frac{m_{P}e^{2}}{2n}\sqrt{\frac{\pi}{\alpha n}}\left(-1+\frac{3}{2n}\right).
 \end{split}
\end{equation}
Now we substitute the form of $n$ to have the complete expression for mass difference, and hence, from eq.(\ref{massspectrum2}), we have\footnote{We have considered upto fourth order of charge contribution in our analysis.}
\begin{equation}
    n\simeq \frac{16\pi M^{2}}{\alpha m_{P}^{2}}\left(1-\frac{\alpha m_{P}^{2}}{32\pi M^{2}}-\frac{e^{2}m_{P}^{4}}{2M^{2}}-\frac{e^{4}m_{P}^{4}}{16M^{4}}\right).
\end{equation}
 After substituting $n$, one can rewrite the mass difference between two consecutive energy states as
\begin{equation}
    \Delta M\simeq \frac{\alpha m_{P}^{2}}{32\pi M}\left(1-\frac{\alpha m_{P}^{2}}{64\pi M^{2}}+\frac{\alpha^{2}m_{P}^{4}}{2\times (32)^{2}\pi^{2}M^{4}}-\frac{e^{4}m_{P}^{4}}{16M^{4}}\right).
\end{equation}
At this stage, if we compare eq.(\ref{massdiff}) and eq.(\ref{massdiff2}), then the mass difference for quantum corrected Schwarzschild and Reissner-Nordstr$\ddot{o}$m black hole are similar in structure.
\noindent Now, we shall compute the corrected form of black hole temperature. Again applying first law of thermodynamics, one can obtain 
    \begin{align}\label{temperature4}
        \mathcal{T}=&\frac{\Delta M}{\Delta S} \nonumber \\ \simeq &\frac{m_{P}^{2}}{8\pi M}\left[ 1-\frac{\alpha m_{P}^{2}}{64\pi M^{2}}+\frac{\alpha^{2}m_{P}^{4}}{2\times (32)^{2}\pi^{2}M^{4}}-\frac{e^{4}m_{P}^{4}}{16M^{4}}\right]
    \end{align}
where $\Delta S=\frac{\alpha}{4}$ has been substituted from the semi-classical entropy spectrum to obtain the temperature. 
Now the Hawking temperature of the Reissner-Nordstr$\ddot{o}$m black hole can be obtained as
\begin{equation}
    T_{H}^{RN}=\frac{f^{\prime}(r_{h})}{4\pi}\simeq \frac{m_{P}^{2}}{8\pi M}\left(1-\frac{e^{4}m_{P}^{4}}{16M^{4}} \right).
\end{equation}
Now in terms of the Hawking temperature, we can rewrite eq.(\ref{temperature4}) as
\begin{equation}
    \mathcal{T} \simeq T_{H}^{RN}\left[1-\frac{\alpha m_{P}^{2}}{64\pi M^{2}}+\frac{\alpha^{2}m_{P}^{4}}{2\times (32)^{2}\pi^{2}M^{4}} \right].
\end{equation}
This equation shows the deviation of the temperature with respect to the Hawking temperature of the black hole.\\
\noindent Now we shall move to calculate the corrected entropy in this non-rotating charged black hole geometry. Remembering the first law of black hole thermodynamics, one can compute
    \begin{align}\label{entropy2}
\mathcal{S} &= \int \frac{dM}{\mathcal{T}}\nonumber \\ &\simeq \frac{4\pi}{m_{P}^{2}}\left[M^{2}+\frac{\alpha m_{P}^{2}}{32\pi}\ln{\left(\frac{M}{m_{P}}\right)}\right. \nonumber \\& ~~~~~~~~~~~~\left.+\frac{m_{P}^{4}}{M^{2}}\left(\frac{\alpha^{2}}{4\times(32)^{2}\pi^{2}}-\frac{e^{4}}{16} \right) \right].
    \end{align}
To conclude about the deviation of entropy, we have to write it down in terms of the area of the black hole. From eq.(\ref{massspectrum2}), we shall put the mass into eq.(\ref{entropy2}) and one can recast the form of entropy as
    \begin{align}
        \mathcal{S}&\simeq \frac{Am_{P}^{2}}{4}+\frac{\alpha}{16}\ln{(A)}+\frac{\alpha}{2Am_{P}^{2}}\left(e^{2}\pi+\frac{\alpha}{32} \right)\nonumber \\ &-\frac{e^{2}\pi \alpha}{A^{2}m_{P}^{4}}\left(e^{2}\pi+\frac{\alpha}{8} \right)+\frac{3\pi^{2}\alpha^{2}e^{4}}{4A^{3}m_{P}^{6}}+2\pi e^{2}+\frac{\alpha}{8}\ln{\left(\frac{m_{P}}{4\sqrt{\pi}}\right)}.
    \end{align}
Here also the leading term follows the usual Bekenstein-Hawking area law, followed by logarithmic deviation and corrections of $\mathcal{O}(\frac{1}{A})$, $\mathcal{O}(\frac{1}{A^{2}})$ and $\mathcal{O}(\frac{1}{A^{3}})$. This expression also matches exactly with the entropy formula computed for the quantum corrected Schwarzschild case.

\section{Mass spectrum, corrected temperature and entropy of Kerr black hole}\label{V}
\noindent In the previous sections, we considered black holes which are spherically symmetric. We now relax the spherical symmetry and consider rotating black holes. Kerr black holes are the axisymmetric vacuum solution of Einstein's field equations. The solution for these black holes was first introduced by Kerr in 1963 \cite{Kerr}. The geometry reads

    \begin{align}\label{kerrmetric}
        ds^{2}&=\frac{\Delta - a^{2}\sin^{2}{\theta}}{\rho^{2}}dt^{2}+\frac{4arG_{0}M}{\rho^{2}}\sin^{2}{\theta}d\phi dt \nonumber \\ & ~~~- \frac{\rho^{2}}{\Delta}dr^{2}-\rho^{2}d\theta^{2}-\frac{\mathcal{A}\sin^{2}{\theta}}{\rho^{2}}d\phi^{2}
    \end{align}
where the parameters are defined as 
\begin{align}\label{parameters}
    \Delta = &r^{2}-2G_{0}Mr+a^{2}\\
    \rho^{2}=&r^{2}+a^{2}\cos^{2}{\theta}\\
    \mathcal{A}=& (r^{2}+a^{2})^{2}-a^{2}\sin^{2}{\theta}~.
\end{align}
The angular momentum of the black hole is defined as $J=aM$. The lapse function for this kind of rotating neutral black hole is
\begin{equation}\label{Kerrlapse}
    f(r) = 1-\frac{2G_{0}Mr}{(r^{2}+a^{2}\cos^{2}{\theta})}~.
\end{equation}
In $a\rightarrow 0$ limit, one gets the Schwarzschild solution.
The solution in eq.(\ref{kerrmetric}) is the Kerr metric in Boyer-Lindquist co-ordinates ($t,r,\theta,\phi$). The most noticable thing in the Kerr metric is that for $\Delta \rightarrow 0$, the coefficient of $dr^{2}$ diverges. Setting $\Delta = 0$ with $a\ne 0$, the radius of the event horizon of Kerr black hole can be obtained \cite{DerekRaine}. There are two horizons for the Kerr black hole. Here, we only consider the event (outer) horizon and it has the form 
\begin{equation}
    r_{h}= G_{0}M+G_{0}M\sqrt{1-\frac{J^{2}}{G_{0}^{2}M^{4}}}~.
\end{equation}
Since, this type of black hole does not contain spherical symmetry, the area is dependent on the rotation parameter $a$ and given by\footnote{Since the angular momentum of the black hole is reasonably fainter compared to the mass, so, we have taken $\frac{J}{M}$ as a small quantity and considered terms upto $\mathcal{O}(J^{2})$ in our analysis.}
\begin{align}
    A=&\sqrt{\mathcal{A}}\int_{0}^{2\pi}d\phi \int_{0}^{\pi}\sin{\theta}~ d\theta =4\pi(r_{h}^{2}+a^{2})\nonumber \\ \simeq &16\pi G_{0}^{2}M^{2}-\frac{4\pi J^{2}}{M^{2}}
\end{align}
where the form of $\mathcal{A}$ is mentioned earlier in eq.(\ref{parameters}). Now we shall do the similar kind of analysis as the previous sections. The mass spectrum for the rotating neutral black hole can be obtained as mentioned in \cite{Bagchi}
\begin{align}\label{massspectrum3}
    M(n)\simeq &\sqrt {\frac{A}{16\pi G_{0}^{2}}+\frac{4\pi J^{2}}{A}}\nonumber \\ \simeq &     \frac{m_{P}}{4}\sqrt{\frac{\alpha (n+\frac{1}{2})}{\pi}}\left[1+\frac{32\pi^{2}J^{2}}{\alpha^{2}(n+\frac{1}{2})^{2}} \right].
\end{align}
Here, the dependence of $n$ is slightly different from the other three cases. If we extract the value of $n$ in terms of $M$ from eq.(\ref{massspectrum3}), we get
\begin{equation}
    n\simeq \frac{16\pi M^{2}}{\alpha m_{P}^{2}}\left(1-\frac{\alpha m_{P}^{2}}{32\pi M^{2}}-\frac{J^{2}m_{P}^{4}}{4M^{4}} \right).
\end{equation}
Now, if the Kerr black hole makes a transition from the $|n+1\rangle$ state to the $|n\rangle$ state, then the mass difference between the two successive energy levels is given by
\begin{align}
    \Delta M \simeq & \frac{m_{P}}{8}\sqrt{\frac{\alpha}{\pi n}}\left ( 1- \frac{1}{2n}+\frac{13}{32n^{2}}\right)\nonumber \\ & ~~~~~~~~~~~-12m_{P}J^{2}\sqrt{\frac{\pi^{3}}{\alpha^{3}n^{5}}}\left( 1-\frac{5}{2n}\right)\nonumber \\
    \simeq & \frac{\alpha m_{P}^{2}}{32\pi M}\left(1-\frac{\alpha m_{P}^{2}}{64\pi M^{2}}+\frac{\alpha^{2}m_{P}^{4}}{2\times (32)^{2}\pi^{2}M^{4}}-\frac{J^{2}m_{P}^{4}}{4M^{4}}\right).
\end{align}
Due to this quantization of mass, the temperature of the black hole will be corrected. This correction can be computed from the first law of thermodynamics as
    \begin{align}\label{Kerrtemp}
        \mathcal{T}= &\frac{\Delta M}{\Delta S}\nonumber \\ \simeq &\frac{ m_{P}^{2}}{8\pi M}\left(1-\frac{\alpha m_{P}^{2}}{64\pi M^{2}}+\frac{\alpha^{2}m_{P}^{4}}{2\times (32)^{2}\pi^{2}M^{4}}-\frac{J^{2}m_{P}^{4}}{4M^{4}}\right).
    \end{align}
This is the corrected temperature obtained from the mass and the entropy spectrum. From the lapse function eq.(\ref{Kerrlapse}), we can calculate the Hawking temperature of the Kerr black hole. In the equatorial plane $\theta = 0$, the form of Hawking temperature is
\begin{equation}\label{Hawkingtempkerr}
    T_{H}^{Kerr}=\frac{f^{\prime}(r_{h})}{4\pi}\simeq \frac{m_{P}^{2}}{8\pi M}\left(1-\frac{J^{2}m_{P}^{4}}{4M^{4}} \right).
\end{equation}
Merging eq.(\ref{Kerrtemp}) and eq.(\ref{Hawkingtempkerr}), we get the deviation of Hawking temperature due to the discrete mass spectrum of the black hole to be
\begin{equation}
    \mathcal{T}\simeq T_{H}^{Kerr}\left[1-\frac{\alpha m_{P}^{2}}{64\pi M^{2}}+\frac{\alpha^{2}m_{P}^{4}}{2\times (32)^{2}\pi^{2}M^{4}} \right].
\end{equation}
\noindent This corrected temperature will now lead to the correction to the black hole entropy. One can write the corrected entropy in terms of the mass of the black hole as following
\begin{align}\label{entropykerr}
    \mathcal{S}&=\int \frac{dM}{\mathcal{T}}\nonumber \\ &
    \simeq  \frac{4\pi}{m_{P}^{2}}\left[M^{2}+\frac{\alpha m_{P}^{2}}{32\pi}\ln{\left(\frac{M}{m_{P}}\right)}\right. \nonumber \\ & ~~~~~~~~~~~~~~~\left.+\frac{m_{P}^{4}}{8M^{2}}\left( \frac{\alpha^{2}}{(32)^{2}\pi^{2}}-J^{2}\right) \right].
\end{align}
By substituting eq.(\ref{massspectrum3}), one can recast eq.(\ref{entropykerr})in terms of the area of the black hole as
    \begin{align}
        \mathcal{S}\simeq \frac{Am_{P}^{2}}{4}&+\frac{\alpha}{16}\ln{\left(A\right)}+\frac{\alpha^{2}}{64Am_{P}^{2}}+\frac{4\pi^{2}\alpha J^{2}}{A^{2}m_{P}^{4}}\nonumber \\ & -\frac{\alpha^{2}\pi^{2}J^{2}}{A^{3}m_{P}^{6}}+\frac{\alpha}{8}\ln{\left(\frac{m_{P}}{4\sqrt{\pi}}\right)}.
    \end{align}
The leading term in the entropy formula is the usual Bekenstein-Hawking entropy followed by the leading order logarithmic correction and subleading inverse area terms.

\section{Conclusion}
\noindent In this work, we have first briefly revisited the connection between black hole thermodynamics and the Landauer's principle on non-rotating neutral spherically symmetric black hole geometries \cite{CortesLiddle}. In the very first section, we have discussed the unification of black-hole energy and Landauer energy and proposed that the energy dissipated due to the Hawking entropy works as the information eraser. From the Landauer's saturation bound, we have computed the value of the semiclassical parameter $\alpha$. Then following the idea in \cite{BekensteinMukhanov} and the recent analysis in \cite{Bagchi}, in the next section, we have computed the mass spectrum of the Schwarzschild black hole from the concept of area quantization of black holes. The mass spectrum of the black hole allows us to calculate the change in mass of the black hole when it makes a transition from one mass state to its consecutive mass state. This leads to the corrected Hawking temperature from the first law of black hole thermodynamics. Applying the first law once again with the corrected Hawking temperature, gives us the corrected entropy of the black hole. We carry out this analysis for the Schwarzschild, quantum improved Schwarzschild, Reissner-Nordstr$\ddot{o}$m and Kerr black holes. The mass spectrum in all these cases are obtained. The corrected temperature and the entropy are also calculated. The corrected entropy is found to contain logarithmic and inverse area corrections which reveals the universality of these corrections \cite{Kaul,DasMajumdar,gup1,gup2,gup3}.

\end{document}